\documentclass[
 reprint,amsmath,amssymb,aps,
]{revtex4-2}

\usepackage{graphicx}
\usepackage{dcolumn}
\usepackage{bm}
\usepackage{hyperref}
\hypersetup{colorlinks, linkcolor={blue}, citecolor={blue}, urlcolor={blue}}

\begin{document}
\title{Electrical Injection and Transport of Coherent Magnons in Non-Collinear Antiferromagnets}

\author{Ping Tang$^{1}$}
\email{tang.ping.a2@tohoku.ac.jp}
\author{Gerrit E. W. Bauer$^{1-3}$}
\affiliation{$^1$WPI-AIMR, Tohoku
University, 2-1-1 Katahira, Sendai 980-8577, Japan}
\affiliation{$^2$Institute for Materials Research and CSIS, Tohoku University,
2-1-1 Katahira, Sendai 980-8577, Japan} 
\affiliation{$^3$Kavli Institute for Theoretical Sciences, University of the Chinese Academy of Sciences,
Beijing 10090, China}
\date{\today}

\begin{abstract}
Non-collinear antiferromagnets (nAFMs) with a small net magnetic moment offer new opportunities for ultrafast spintronic devices, owing to unique physical properties. While in ferromagnets and collinear AFMs the spin current polarization is locked to the magnetization $\hat{\mathbf{m}}$ and N\'eel vector $\hat{\mathbf{n}}$ directions, we predict that magnon spin currents injected by metal contacts into nAFMs can be polarized with both $\hat{\mathbf{n}}$ and $\hat{\mathbf{m}}$ components when carried by a coherent superposition of magnon eigenstates. The spin injection efficiency is governed by an interface spin conductance tensor that depends on the non-collinear magnetic texture. While the $\hat{\mathbf{m}}$-component diffuses freely into the nAFM, the $\hat{\mathbf{n}}$-component oscillates as a function of distance from the injector and applied magnetic field, analogous to the Hanle effect of electron spins in metals. Our findings reveal the potential of nAFMs as platforms for the study of tensorial coherent spin transport.
\end{abstract}

\maketitle


\affiliation{$^1$WPI-AIMR, Tohoku
University, 2-1-1 Katahira, Sendai 980-8577, Japan} 
\affiliation{$^2$Institute for Materials Research and CSIS, Tohoku University,
2-1-1 Katahira, Sendai 980-8577, Japan} 
\affiliation{$^3$Kavli Institute for Theoretical Sciences, University of the Chinese Academy of Sciences,
Beijing 10090, China}
Coherent injection, transport, and manipulation of spins are fundamental challenges in spintronics \cite{RevModPhys.76.323}. AFMs are promising materials for ultrafast and robust spintronics devices due to their intrinsic terahertz spin dynamics and the absence of crosstalk by dipolar couplings \cite{jungwirth2016antiferromagnetic,RevModPhys.90.015005,jungwirth2018multiple,jungfleisch2018perspectives}. In a collinear easy-axis AFM, the quanta of elementary magnetic excitations (magnons) are right- and left-handed precessions of the N\'eel vector with degenerate band dispersions but opposite spin angular momenta of $\pm\hbar$. A spin accumulation in an adjacent metallic contact polarized parallel to the N\'eel vector injects non-equilibrium distributions of both magnon species independently \cite{PhysRevB.94.054413, PhysRevB.95.144408}, leading to purely diffusive spin transport in AFMs that can be described by a two-channel resistor model~\cite{PhysRevB.93.054412, lebrun2018tunable,PhysRevB.99.180405,PhysRevX.9.011026}. We call this injection mechanism \textit{incoherent} since the phases of the two magnon modes are uncorrelated. 

Recent research on spin transport in easy-plane AFMs has opened up new perspectives on coherent magnon transport by means of electrical injection \cite{han2023coherent}. In contrast to easy-axis AFMs, easy-plane anisotropy lifts the degeneracy of two magnon eigenmodes that correspond to the N\'eel vector oscillating in and out of the easy plane. These linearly polarized eigenmodes do not carry net spin angular momentum, a net spin may be transported by their coherent superposition \cite{han2020birefringence,lebrun2020long}. An injected spin current then induces a magnon spin accumulation that exhibits spatial oscillations governed by magnon interband splitting \cite{PhysRevLett.125.247204,PhysRevB.102.174445,PhysRevLett.130.216703,PhysRevB.107.184404,PhysRevB.110.L140408,su2024controllable}. This behavior can be captured by a generalized diffusion equation for magnon pseudospins that \emph{coherently} precess around the perpendicular pseudo-field induced by the easy-plane anisotropy \cite{PhysRevLett.125.247204,PhysRevB.102.174445,PhysRevB.110.L140408}, analogous to the Hanle effect of electronic spin accumulation in metals under a transverse magnetic field \cite{kikkawa1999lateral,jedema2002electrical,fabian2007semiconductor}. However, the microscopic mechanism underlying the injection of coherent magnon superposition states on top of a thermally excited (incoherent) magnon cloud at easy-plane AFM interfaces remains elusive. Existing theoretical models \cite{PhysRevB.102.174445,PhysRevB.110.L140408} assume spin injection into circularly polarized magnon states (i.e., a superposition of linearly polarized eigenmodes), as in collinear easy-axis AFMs. The validity of this assumption, as well as the  physical principles behind the magnon Hanle effect in \emph{non-collinear} AFMs (nAFMs), remain unknown.

nAFMs with a small net magnetic moment garnered considerable attention since they offer new opportunities for antiferromagnetic spintronics \cite{rimmler2024non}. Their non-collinear spin texture can be driven by applied magnetic fields or internal magnetic couplings such as the Dzyaloshinskii-Moriya interaction (DMI). Unlike collinear AFMs, the nAFM ordering breaks spin rotation symmetry without a common quantization axis, implying that the magnon spin may have an arbitrary magnitude and direction, with related rich spin transport physics. Okuma \cite{PhysRevLett.119.107205} reported a momentum-space spin texture of magnons in a two-dimensional kagome nAFM, while we formulated spin pumping and spin Seebeck effects in a two-sublattice nAFM that depend on the degree of non-collinearity and the interfaces to metallic contacts~\cite{PhysRevLett.133.036701}. 

\begin{figure}
    \centering
    \includegraphics[width=8.6 cm]{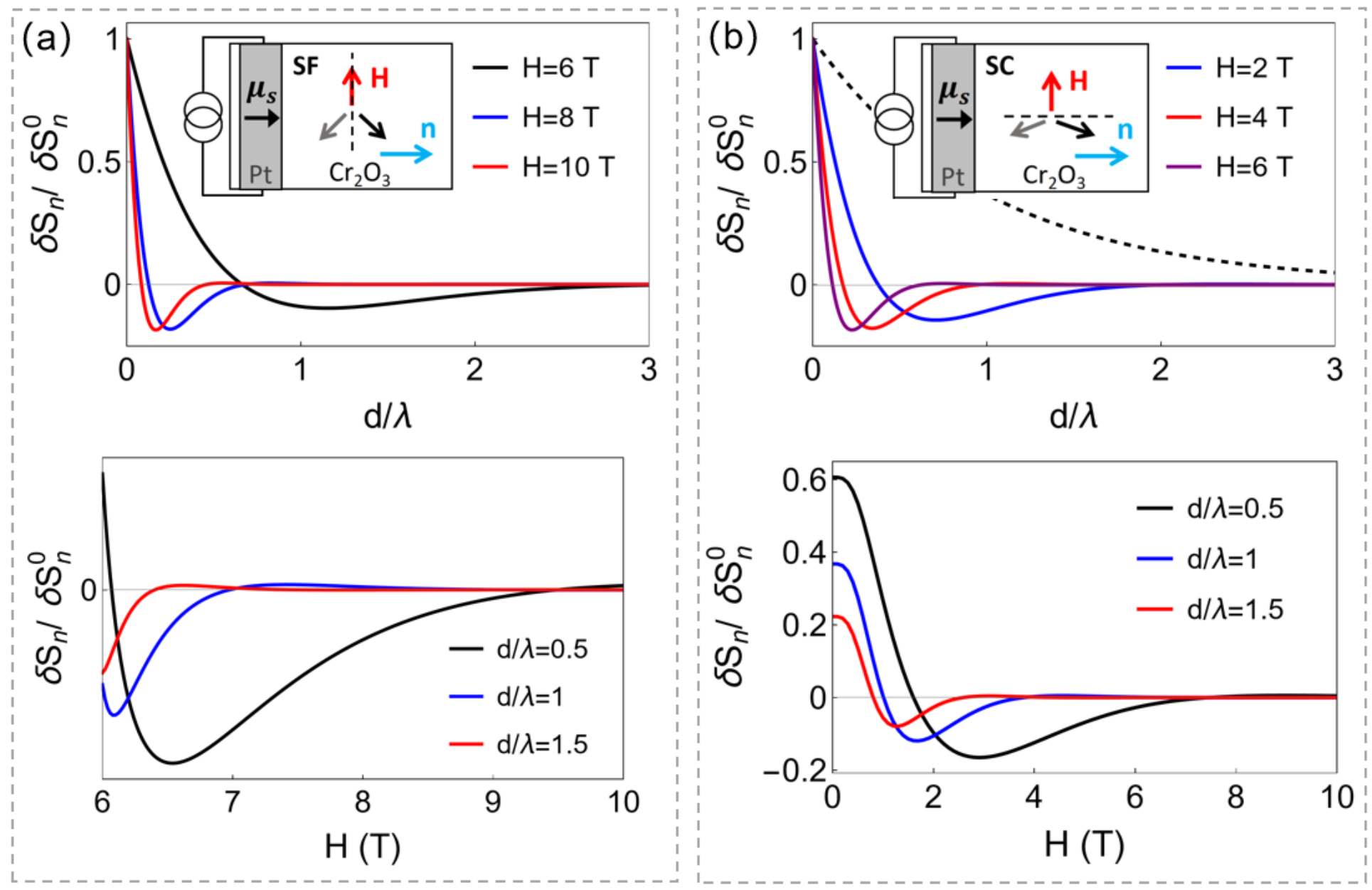}
    \caption{The magnon spin density component \(\delta S_n\) along the N\'eel vector ($\hat{\mathbf{n}}$) in (a) the spin-flop (transition at \(H_{\text{sf}}=5.85\,\)T) and (b) spin-canted states of Cr$_2$O$_3$, injected at room temperature by a spin accumulation $\boldsymbol{\mu}_{s}\parallel\hat{\mathbf{n}}$ in the HM (Pt). Here $\delta S_{n}$ is normalized by its value at the interface, and \(d/\lambda\) is the distance from the interface divided by the nominal magnon diffusion length $\lambda=\sqrt{D_{\alpha\beta}\tau_{D}}$. The dashed curve in (b) is for the collinear state without an applied field. The parameters for Cr$_2$O$_3$ are $H_{E}=245\,$T, $H_{\parallel}=0.07\,$T, $H_{\perp}=0\,$T, and $H_{D}=0\,$T \cite{PhysRev.130.183}.}
    \label{Fig-transport}
\end{figure}

In this Letter, we present a microscopic theory of spin injection and transport in a generic two-sublattice nAFM in the spin-flop and spin-canted phases. The former emerges in the presence of a sufficiently strong magnetic field along the N\'eel vector of an easy-axis collinear AFM, while the latter is caused by the DMI and/or a magnetic field normal to the N\'eel vector. A spin accumulation in a metal contact can inject a magnon spin current that is polarized with both $\hat{\mathbf{n}}$ and $\hat{\mathbf{m}}$ components and carried by a coherent superposition of the magnon eigenmodes of the bulk nAFM. The spin injection efficiency varies strongly with the geometry and the canting angles of two sublattice magnetizations, consistent with the configuration dependence of non-local spin transport in hematite \cite{lebrun2018tunable}. We predict that the non-collinear spin order causes a magnon Hanle effect with an additional phase contribution from the injector not seen in collinear AFMs \cite{PhysRevB.102.174445}. We illustrate our results by predicting an experimentally observable field-controlled magnon Hanle effect in the non-collinear states of the easy-axis AFM Cr$_2$O$_3$ [see Fig.~\ref{Fig-transport}], which is absent in the parent collinear phase.

 \begin{figure}
    \centering
    \includegraphics[width=8.6cm]{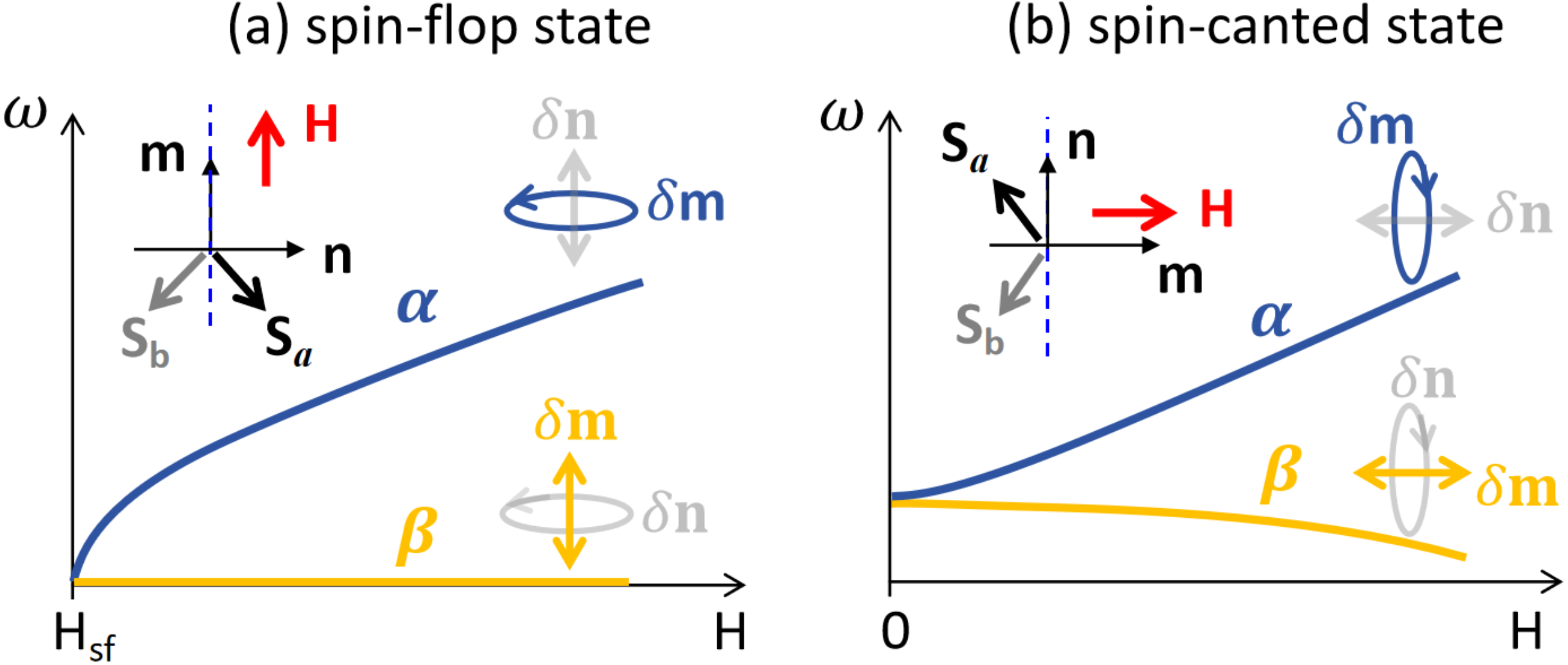}
    \caption{The magnon eigenmodes and associated magnetic dynamics in the spin-flop and spin-canted states of two-sublattice AFMs, with the blue dashed line indicating the easy axis. These two nAFM states in the main text are uniformly described in a coordinate system spanned by the orthogonal equilibrium N\'eel vector ($\mathbf{n}$) and net magnetization ($\mathbf{m}$).}
    \label{Fig-nAFM}
\end{figure}

We focus on the non-collinear state of a generic two-sublattice AFM, including the spin-canted state, e.g., in hematite, induced by the DMI and/or a transverse magnetic field, and the spin-flop state driven by a sufficiently large magnetic field along the
easy axis. Applying the standard Holstein-Primakoff (HP) transformation to the spin operators of each sublattice leads to a quadratic magnon Hamiltonian in the form \cite{rezende2020fundamentals,PhysRevLett.133.036701}
\begin{align}
\hat{\mathcal{H}}=  &\sum_{\mathbf{q}}\mathcal{A}_{\mathbf{q}}(\hat
{a}_{\mathbf{q}}^{\dagger}\hat{a}_{\mathbf{q}}+\hat{b}_{\mathbf{q}}^{\dagger
}\hat{b}_{\mathbf{q}})+\mathcal{B}_{\mathbf{q}}(\hat{a}_{\mathbf{q}}\hat
{b}_{-\mathbf{q}}+\text{H.c.})\nonumber\\
&  +\frac{1}{2}\mathcal{C}_{\mathbf{q}}(\hat{a}_{\mathbf{q}}\hat
{a}_{-\mathbf{q}}+\hat{b}_{\mathbf{q}}\hat{b}_{-\mathbf{q}}+\text{H.c.}%
)+\mathcal{D}_{\mathbf{q}}(\hat{a}_{\mathbf{q}}\hat{b}_{\mathbf{q}}^{\dagger
}+\text{H.c.})\nonumber\\
=  &  \hbar\sum_{\mathbf{q}}(\omega_{\mathbf{q}\alpha}\hat{\alpha}%
_{\mathbf{q}}^{\dagger}\hat{\alpha}_{\mathbf{q}}+\omega_{\mathbf{q}\beta}%
\hat{\beta}_{\mathbf{q}}^{\dagger}\hat{\beta}_{\mathbf{q}})+\text{const.}
\label{mHM}%
\end{align}
where $\mathcal{A}_{\mathbf{q}}=H_{E}\cos2\theta+ H\sin
\theta+\frac{H_{\parallel}}{2}[3\cos^{2}\theta_{A}-1]+\frac{H_{\perp}}{2}+H_{D}\sin2\theta$,
$\mathcal{B}_{\mathbf{q}}=H_{E}\mathcal{F}_{\mathbf{q}}\cos^{2}%
\theta+\frac{H_{D}}{2}\mathcal{F}_{\mathbf{q}}\sin2\theta$, $\mathcal{C}%
_{\mathbf{q}}=\frac{H_{\parallel}}{2}\sin^{2}\theta_{A}+\frac{H_{\perp}}{2}$, and $\mathcal{D}%
_{\mathbf{q}}=H_{E}\mathcal{F}_{\mathbf{q}}\sin^{2}\theta-\frac{H_{D}}{2}
\mathcal{F}_{\mathbf{q}}\sin2\theta$, and $\mathcal{F}_{\mathbf{q}}=(1/z_{0}%
)\sum_{\boldsymbol{\delta}_{ij}}\exp(i\mathbf{q}\cdot\boldsymbol{\delta}%
_{ij})$ the form factor with coordination number $z_{0}$. $H_{E}$, $H_{\parallel}$, $H_{\perp}$, $H_{D}$ and $H$ are exchange, easy-axis anisotropy, easy-plane anisotropy, DMI, and applied fields, respectively. $\theta$ and $\theta_{A}$ are the canting angles of the equilibrium
relative to the N\'eel vector and easy axis, respectively, i.e., $\theta
=\arcsin[(H+H_{D})/(2H_{E}+H_{\parallel})]$ and $\theta_{A}=\theta$ for (i) and $\theta=\arcsin[(
H+H_{D})/(2H_{E}-H_{\parallel})]$ and $\theta_{A}=\pi/2-\theta$ and $\theta_{A}=\theta$ for (ii). The dispersion of the two magnon branches
$\hbar\omega_{\mathbf{q}\alpha}=\sqrt{(\mathcal{A}_{\mathbf{q}}+\mathcal{D}%
_{\mathbf{q}})^{2}-(\mathcal{C}_{\mathbf{q}}+\mathcal{B}_{\mathbf{q}})^{2}}$
and $\hbar\omega_{\mathbf{q}\beta}=\sqrt{(\mathcal{A}_{\mathbf{q}}%
-\mathcal{D}_{\mathbf{q}})^{2}-(\mathcal{C}_{\mathbf{q}}-\mathcal{B}%
_{\mathbf{q}})^{2}}$ follows from the Bogoliubov-de Gennes transformation
\begin{equation}
\left(
\begin{matrix}
\hat{a}_{\mathbf{q}}\\
\hat{b}_{\mathbf{q}}\\
\hat{a}_{-\mathbf{q}}^{\dagger}\\
\hat{b}_{-\mathbf{q}}^{\dagger}%
\end{matrix}
\right)  =\left(
\begin{matrix}
Q_{11} & Q_{12} & Q_{13} & Q_{14}\\
Q_{11} & -Q_{12} & Q_{13} & -Q_{14}\\
Q_{13} & Q_{14} & Q_{11} & Q_{12}\\
Q_{13} & -Q_{14} & Q_{11} & -Q_{12}%
\end{matrix}
\right)  \left(
\begin{matrix}
\hat{\alpha}_{\mathbf{q}}\\
\hat{\beta}_{\mathbf{q}}\\
\hat{\alpha}_{-\mathbf{q}}^{\dagger}\\
\hat{\beta}_{-\mathbf{q}}^{\dagger}%
\end{matrix}
\right)  \label{Transform}%
\end{equation}
with real coefficients:
\begin{align}
Q_{11}=  &  \sqrt{\frac{\mathcal{A}_{\mathbf{q}}+\mathcal{D}_{\mathbf{q}%
}+\hbar\omega_{\mathbf{q}\alpha}}{4\hbar\omega_{\mathbf{q}\alpha}}}%
,Q_{12}=\sqrt{\frac{\mathcal{A}_{\mathbf{q}}-\mathcal{D}_{\mathbf{q}}%
+\hbar\omega_{\mathbf{q}\beta}}{4\hbar\omega_{\mathbf{q}\beta}}},\nonumber\\
Q_{13}=  &  -\mathrm{sgn}(\mathcal{C}_{\mathbf{q}}+\mathcal{B}_{\mathbf{q}%
})\sqrt{\frac{\mathcal{A}_{\mathbf{q}}+\mathcal{D}_{\mathbf{q}}-\hbar
\omega_{\mathbf{q}\alpha}}{4\hbar\omega_{\mathbf{q}\alpha}}},\nonumber\\
Q_{14}= &  -\mathrm{sgn}(\mathcal{C}_{\mathbf{q}}-\mathcal{B}_{\mathbf{q}%
})\sqrt{\frac{\mathcal{A}_{\mathbf{q}}-\mathcal{D}_{\mathbf{q}}-\hbar
\omega_{\mathbf{q}\beta}}{4\hbar\omega_{\mathbf{q}\beta}}}.
\end{align}

The direction and magnitude of the angular momentum of a single magnon are essential quantities in spintronics. In collinear AFMs with axial rotation symmetry, the two branches of magnons carry spin angular momenta $\pm\hbar$ along the N\'eel vector. In nAFMs, the magnon spin is no longer quantized due to the breaking of spin-rotation symmetry. By expanding the spin operators in terms of the eigenstate basis ($\hat{\alpha}_{\mathbf{q}},\hat{\beta}_{\mathbf{q}}$), the magnon spin is expressed as a non-diagonal matrix 
\begin{align}
\hat{\mathbf{s}}&=\left( \begin{matrix}
 s_{\alpha} &0\\
 0&s_{\beta}
\end{matrix} \right)\hat{\mathbf{m}}+\left(
\begin{matrix}
0, & s_{\alpha\beta} \\
s_{\alpha\beta}^{\ast}&  0
\end{matrix}
\right)\hat{\mathbf{n}} \label{magnonspin}
\end{align}    
where $s_{\alpha}=2\hbar\sin\theta\vert Q_{11}\vert^2+\vert Q_{13}\vert^{2}$ and $s_{\beta}=2\hbar\sin\theta\vert Q_{12}\vert^2+\vert Q_{14}\vert^2$ are the spins of the magnon eigenmodes parallel to $\hat{\mathbf{m}}$, while $s_{\alpha\beta}=-2\hbar\cos\theta (Q_{11}Q_{12}+Q_{13} Q_{14})$ is the spin component of a coherent superposition state along $\hat{\mathbf{n}}$, where $\hat{\mathbf{m}}$ and $\hat{\mathbf{n}}$ denote the unit directions of the equilibrium net magnetization and N\'eel vector, respectively. 

 \begin{figure}
    \centering
    \includegraphics[width=5.6 cm]{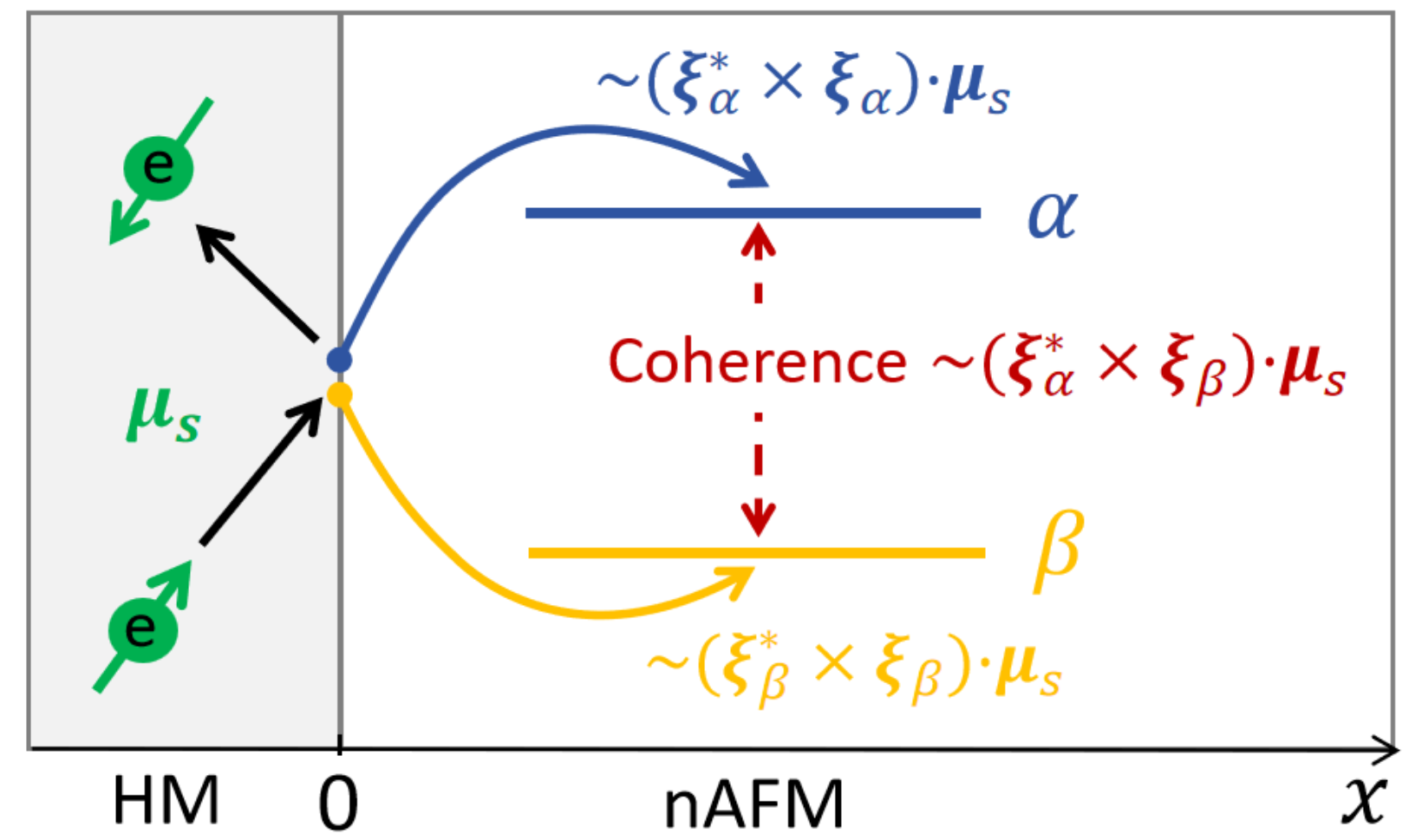}
    \caption{Injection of interband-coherent magnons into a two-sublattice nAFM by a spin-flip scattering at the interface. The cross-products of the associated complex-valued coupling vectors defined by Eq.~(\ref{SdH}) govern the injection rate. We recover results for a collinear easy-plane AFM with $\boldsymbol{\xi}_{\alpha}\sim \hat{\mathbf{m}}\times\hat{\mathbf{n}}$, $\boldsymbol{\xi}_{\beta}\sim i\hat{\mathbf{m}}$, into which $\boldsymbol{\mu}_{s}\parallel\hat{\mathbf{n}}$ injects a coherent superposition of two linearly polarized eigenmodes.}
    \label{Fig-scatter}
\end{figure}

We now address the transport of spins in the nAFM injected by a heavy metal (HM) contact at the interface $x=0$ [Fig.~\ref{Fig-scatter}]. In the case of collinear AFMs, spin injection requires a spin accumulation ($\boldsymbol{\mu}_{s}$) component parallel to $\hat{\mathbf{n}}$, while a perpendicular component is absorbed as a spin transfer torque. The spin injection into two magnon bands is then uncoupled and uncorrelated. We show below that when the magnetic texture is non-collinear, the spin accumulation in the HM may excite a superposition of magnon eigenstates, leading to a magnon spin current with polarization components along both $\hat{\mathbf{n}}$ and $\hat{\mathbf{m}}$. An injected spin component along $\hat{\mathbf{m}}$ freely diffuses into the bulk of the nAFM in the form of non-equilibrium magnons in both bands, while the $\hat{\mathbf{n}}$-component from a coherent superposition of the two magnons is accompanied by a Larmor-like oscillation. Both processes can be captured by the kinetic equation of the spatiotemporal $2\times 2$ density matrix $\hat{\rho}(x,\mathbf{q}, t)=\left(\begin{matrix}
  \rho_{\alpha} & \rho_{\alpha\beta}\\
  \rho_{\alpha\beta}^{\ast} & \rho_{\beta}
\end{matrix}\right)$: 
\begin{equation}
\mathcal{L}(\hat{\rho})\equiv\frac{\partial \hat{\rho}}{\partial t}+\frac{1}{2}\left\{\hat{v}_{x},\nabla_{x}\hat{\rho}\right\}+\frac{i}{\hbar}[\hat{\mathcal{H}},\hat{\rho}]=\left.\frac{\partial \hat{\rho}}{\partial t}\right\vert_{\text{rel}}+\delta(x)\hat{I}
\label{Kinet}
\end{equation}
where $\mathcal{L}(\hat{\rho})$ represents the kinetic operator on the l.h.s., $\hat{v}_{x}=\partial\hat{\mathcal{H}}/(\hbar\partial q_{x})$ the $x$-component of the group velocity, $\{\cdots\}$ is an anticommutator, and $\hat{\mathcal{H}}=\text{diag}(\hbar\omega_{\mathbf{q}\alpha},\hbar\omega_{\mathbf{q}\beta})$ the magnon Hamiltonian in the eigenmode basis. The diagonal elements of $\hat{\rho}$ represent the probabilities to find magnons in the corresponding eigenmodes, while the off-diagonal describe the interband coherence. In Eq.~(\ref{Kinet}), the commutator $[\hat{\mathcal{H}},\hat{\rho}]$ captures the internal coherent magnon dynamics, while $(\partial\hat{\rho}/\partial t)\vert_{\text{rel}}$ on the r.h.s. incorporates incoherent scattering by impurities, phonons, and other magnons. In the relaxation time approximation \cite{PhysRevLett.109.096603}
\begin{equation}
\left.\frac{\partial \hat{\rho}}{\partial t}\right\vert_{\text{rel}} =-\frac{\hat{\rho}-\bar{\hat{\rho}}}{\tau_{0}}-\frac{\hat{\rho}-\hat{\rho}^{(eq)}}{\tau_{D}} \label{Rel}
\end{equation}
where $\tau_{0}$ and $\tau_{D}$ are the phenomenological relaxation times for magnon-conserving and -non-conserving processes that relax $\hat{\rho}$ toward $\bar{\hat{\rho}}$ and $\hat{\rho}^{(eq)}$, respectively. Here $\bar{\hat{\rho}}$ is the average of $\hat{\rho}$ over the direction of $\mathbf{q}$~\cite{Noterho}, while $\hat{\rho}^{(eq)}=\text{diag}(n_{\mathbf{q}\alpha}, n_{\mathbf{q}\beta})$ is the temperature-dependent equilibrium density matrix. 

The injection matrix $\hat{I}$ on the r.h.s. of Eq.~(\ref{Kinet}) is a rate of change caused by the spin injection at the interface $x=0$. Here we adopt the s-d exchange interaction between local spins in the nAFM and conduction electrons in the HM, which gives magnon-electron scattering \cite{PhysRevLett.133.036701}:
\begin{align}
\hat{\mathcal{H}}_{sd}=  &  V_{sd}\sum_{\mathbf{k}\mathbf{k}^{\prime
}\mathbf{q}}\hat{c}_{\mathbf{k}s^{\prime}}^{\dagger}\hat{c}_{\mathbf{k}%
^{\prime}s}\left( \hat{\alpha}_{\mathbf{q}}^{\dagger} \boldsymbol{\xi}_{\alpha}+\hat{\beta}_{\mathbf{q}}^{\dagger}\boldsymbol{\xi}_{\beta}\right)\cdot
\boldsymbol{\sigma}_{s^{\prime}s}+\mathrm{H.c.} \label{SdH}%
\end{align}
where $V_{sd}\simeq(J_{a}+J_{b})\sqrt{S/(2N_{m})}$ is a point scattering
potential, with $J_{a}$ and $J_{b}$ are the s-d exchange integrals for two
sublattices with local spins $S$, and $N_{m}$ is the number of magnetic unit cells in the nAFM. $\hat{c}_{\mathbf{k}s}^{(\dagger)}$ is the annihilation (creation) operator of electrons with spin index $s$ and wave vector $\mathbf{k}$, and $\boldsymbol{\sigma
}_{s^{\prime}s}$ represents an element of the vector of Pauli matrices, with the Einstein summation convention implied over repeated spin indices here and hereafter. The complex-valued coupling vectors $\boldsymbol{\xi}_{\alpha}$ and $\boldsymbol{\xi}_{\beta}$ are associated with the dynamics of the average nAFM spins at the interface: $\boldsymbol{\xi}_{\alpha}=(Q_{11}+Q_{13})\hat{\mathbf{m}}\times\hat{\mathbf{n}}+i\varrho_{u}\cos\theta(Q_{11}%
-Q_{13})\hat{\mathbf{m}}+i\sin\theta(Q_{11}-Q_{13})\hat{\mathbf{\mathbf{n}}}$,
 $\boldsymbol{\xi}_{\beta}=\varrho_{u}(Q_{12}+Q_{14})\hat{\mathbf{m}}
\times\hat{\mathbf{n}}+i\cos\theta(Q_{12}%
-Q_{14})\hat{\mathbf{m}}+i\varrho_{u}\sin\theta(Q_{12}-Q_{14})\hat{\mathbf{n}}$, where $\varrho_{u}\equiv(J_{a}%
-J_{b})/(J_{a}+J_{b})$ parameterizes the magnetic compensation at the interface \cite{Notevarrho}. For a fully compensated interface with $\varrho_{u}=0$, $\boldsymbol{\xi}_{\alpha}=(Q_{11}+Q_{13})\hat{\mathbf{m}}\times\hat{\mathbf{n}}+i\sin\theta(Q_{11}-Q_{13})\hat{\mathbf{\mathbf{n}}}$,
 $\boldsymbol{\xi}_{\beta}=i\cos\theta(Q_{12}%
-Q_{14})\hat{\mathbf{m}}$ are then governed by the net magnetization dynamics $\delta\mathbf{m}(t)$ of the corresponding bulk magnon eigenmodes [see Fig.~\ref{Fig-nAFM}]. According to the Fermi’s Golden Rule, Eq.~(\ref{SdH}) leads to  
\begin{align}
I_{\nu\nu^{\prime}}=&\frac{2\pi V_{sd}^{2}\mathcal{V}_{m}}{\hbar A_{I}}\sum_{\mathbf{k}\mathbf{k}^{\prime}}\left( \boldsymbol{\xi}_{\nu}^{\ast}\cdot\boldsymbol{\sigma}_{ss^{\prime}}\right)\left( \boldsymbol{\xi}_{\nu^{\prime}}\cdot\boldsymbol{\sigma}_{s^{\prime}s}\right)\left[f_{\mathbf{k}^{\prime}s}(1-f_{\mathbf{k}s^{\prime}})\right.\nonumber\\
&\left.+n_{%
\mathbf{q}\nu} (f_{\mathbf{k}^{\prime}s}-f_{\mathbf{k}s^{\prime}})\right]\delta(%
\varepsilon_{\mathbf{k}^{\prime}s}-\varepsilon_{\mathbf{k}%
s^{\prime}}-\hbar\omega_{\mathbf{q}\nu} ) \label{I}
\end{align}
where $A_{I}$ and $\mathcal{V}_{m}$ denote the interface area and the volume of the nAFM, respectively. $f_{\mathbf{k}s}$ represent the Fermi-Dirac distribution of electrons with spin index $s$ and energy $\varepsilon_{\mathbf{k}s}$ in the HM, with the spin quantization axis along $\boldsymbol{\mu}_{s}$, while to leading order the magnons are assumed to be in thermal equilibrium \cite{NoteN}. In linear response to $\mu_{s}$ Eq.~(\ref{I}) reduces to 
\begin{align}
I_{\nu\nu^{\prime}}=-\frac{i\pi V_{sd}^{2}\mathcal{N}_{F}^{2}\mathcal{V}_{m}}{\hbar A_{I}}\Gamma_{\nu\nu^{\prime}}(\boldsymbol{\xi}_{\nu}^{\ast}\times\boldsymbol{\xi}_{\nu^{\prime}})\cdot\boldsymbol{\mu}_{s} \label{I2}
\end{align}   
with $I_{\beta\alpha}=I_{\alpha\beta}^{\ast}$. $\mathcal{N}_{F}$ is the density of state of electrons at the Fermi level and $\Gamma_{\nu\nu^{\prime}}=\frac{1}{2}\left[\omega_{\mathbf{q}\nu}\frac{\partial n_{\mathbf{q}\nu}}{\partial\omega_{\mathbf{q}\nu}}+\omega_{\mathbf{q}\nu^{\prime}}\frac{\partial n_{\mathbf{q}\nu^{\prime}}}{\partial\omega_{\mathbf{q}\nu^{\prime}}}\right]$ a temperature-dependent factor. The nonvanishing off-diagonal elements $\sim(\boldsymbol{\xi}_{\alpha}^{\ast}\times\boldsymbol{\xi}_{\beta})\cdot\boldsymbol{\mu}_{s}$ represent the injection of coherent magnons, which source the magnon Hanle effect in the nAFM, as discussed below.

We solve Eq.~(\ref{Kinet}) by invoking efficient magnon-conserving scatterings $\tau_{0}\ll \tau_{D}$ \cite{PhysRevLett.109.096603, PhysRevB.94.014412}, such that $\hat{\rho}$ on the l.h.s. of Eq.~(\ref{Kinet}) can be replaced by $\bar{\hat{\rho}}$. Substituting the ansatz $\hat{\rho}=\bar{\hat{\rho}}-\tau_{0}\mathcal{L}(\bar{\hat{\rho}})$ and averaging both sides over the $\mathbf{q}$ direction for isotropic magnon bands ($\omega_{\mathbf{q}\nu}=\omega_{q\nu}$), we arrive at the coherent magnon diffusion equations:
\begin{align}
&\left(\partial_{t}-D_{\nu}\nabla_{x}^{2}+\tau_{D}^{-1}\right)\bar{\rho}_{\nu}=\frac{\rho_{\nu}^{(eq)}}{\tau_{D}}+\delta(x)I_{\nu}\label{1}\\
&\left\{\partial_{t}-D_{\alpha\beta}\nabla_{x}^{2}+\frac{i\omega_{s}}{1+i\omega_{s}\tau_{0}}+\tau_{D}^{-1}\right\}\bar{\rho}_{\alpha\beta}=\delta(x)I_{\alpha\beta} \label{2}
\end{align}
where $\omega_{s}(q)=\omega_{q\alpha}-\omega_{q\beta}$ are the interband splitting, $D_{\alpha(\beta)}(q)=\tau_{0}v_{\alpha(\beta)}^{2}/3$ and $D_{\alpha\beta}(q)=\tau_{0}(v_{\alpha}+v_{\beta})^{2}/12$ are $q$-dependent diffusion coefficients, and $v_{\alpha(\beta)}=\partial\omega_{q\alpha(\beta)}/\partial q$. For the boundary condition $\bar{\hat{\rho}}\rightarrow \hat{\rho}^{(eq)}$ far from the interface $x\rightarrow \infty$, the steady-state solution of Eqs.~(\ref{1}) and (\ref{2}) reads
\begin{align}\label{rhod}
\bar{\rho}_{\nu}=\rho_{\nu}^{(eq)}+\frac{\lambda_{\nu}I_{\nu}}{D_{\nu}}e^{-x/\lambda_{\nu}},\,\,\,
\bar{\rho}_{\alpha\beta}=\frac{\widetilde{\lambda}_{\alpha\beta}I_{\alpha\beta}}{D_{\alpha\beta}}e^{-x/\widetilde{\lambda}_{\alpha\beta}}
\end{align}
where $\lambda_{\nu}=\sqrt{D_{\nu}\tau_{D}}$ is the mode-dependent diffusion length. The complex parameter $\widetilde{\lambda}_{\alpha\beta}=\lambda_{\alpha\beta}/\sqrt{1+2ik_{s}\lambda_{\alpha\beta}}$ describes the precession and decay of the interband coherence,
\begin{align}
\lambda_{\alpha\beta}=\sqrt{D_{\alpha\beta}}\left\{ \tau_{D}^{-1}+\frac{\omega_{s}^{2}\tau_0}{1+\omega_{s}^{2}\tau_{0}^2}\right\}^{-1/2}
\end{align}
is the transport diffusion length, and $k_{s}=\omega_{s}\lambda_{\alpha\beta}/[2D_{\alpha\beta}(1+\omega_{s}^{2}\tau_{0}^{2})]$ a characteristic wave number. According to Eq.~(\ref{rhod}) $\bar{\rho}_{\nu}$ decays exponentially in space, while the $k_{s}$-dependent oscillations of $\bar{\rho}_{\alpha\beta}$ reflect the coherent dynamics referred to as the magnon Hanle effect. The interband splitting \(\omega_{s}\) suppresses the decay length $\lambda_{\alpha\beta}$ and the oscillation period via $k_{s}$. The non-equilibrium magnon spin accumulation reads
\begin{align}
\delta\boldsymbol{S}=&\int\frac{d\mathbf{q}}{(2\pi)^{3}}\text{Tr}[(\hat{\rho}-\hat{\rho}^{(eq)})\hat{\mathbf{s}}]\nonumber\\
=&\int\frac{d^{3}q}{(2\pi)^{3}}\sum_{\nu} \frac{s_{\nu}\lambda_{\nu}I_{\nu}}{D_{\nu}}  e^{-\frac{x}{\lambda_{\nu}}}\hat{\mathbf{m}}+\int\frac{d^{3}q}{(2\pi)^{3}}\frac{2s_{\alpha\beta}\vert I_{\alpha\beta}\vert}{D_{\alpha\beta} }\hat{\mathbf{n}}\nonumber\\
&\cdot\begin{cases} 
 \frac{e^{-k_{s}^{\prime}x}}{\sqrt{2}k_{s}^{\prime}}\cos\left(k_{s}^{\prime}x-\phi_{I}+\frac{\pi}{4}\right) , & k_{s}\lambda_{\alpha\beta}\gg 1 \\
 \lambda_{\alpha\beta}e^{-\frac{x}{\lambda_{\alpha\beta}}}\cos\left(k_{s}x-\phi_{I}\right) , & k_{s}\lambda_{\alpha\beta}\ll 1
\end{cases}\label{Sm}
\end{align}   
where $\phi_{I}=\arg I_{\alpha\beta}$ and $k_{s}^{\prime}=\sqrt{k_{s}/\lambda_{\alpha\beta}}$. In the limit $k_{s}\lambda_{\alpha\beta}\gg 1$, there is an additional constant phase from $\arg\widetilde{\lambda}_{\alpha\beta}\rightarrow \pi/4$. For a compensated interface ($\varrho_{u}=0$)
\begin{equation}
 \phi_{I}=\tan^{-1}\left\{\sin\theta\frac{  (Q_{11}-Q_{13})(\hat{\mathbf{m}}\times\hat{\mathbf{n}})\cdot\boldsymbol{\mu}_{s}}{ (Q_{11}+Q_{13})(\hat{\mathbf{n}}\cdot\boldsymbol{\mu}_{s})}  \right\} 
\end{equation}
depends on the $\boldsymbol{\mu}_{s}$ direction and vanishes identically in the collinear limit  $\theta\rightarrow 0$. The associated magnon spin current
 \begin{align}
\mathbf{J}_{s}=&\frac{1}{2}\int\frac{d\mathbf{q}}{(2\pi)^{3}}\text{Tr}[\hat{\rho}(\hat{v}_{x}\hat{\mathbf{s}}+\hat{\mathbf{s}}\hat{v}_{x})]\nonumber\\
=&\int\frac{d^{3}q}{(2\pi)^{3}}\sum_{\nu} s_{\nu} I_{\nu}e^{-\frac{x}{\lambda_{\nu}}}\hat{\mathbf{m}}+\int\frac{d^{3}q}{(2\pi)^{3}}2s_{\alpha\beta}\vert I_{\alpha\beta}\vert\hat{\mathbf{n}}\nonumber\\
&\cdot\begin{cases} 
  e^{-k_{s}^{\prime}x}\cos\left(k_{s}^{\prime}x-\phi_{I}\right) , & k_{s}\lambda_{\alpha\beta}\gg 1 \\
 e^{-\frac{x}{\lambda_{\alpha\beta}}}\cos\left(k_{s}x-\phi_{I}\right) , & k_{s}\lambda_{\alpha\beta}\ll 1
\end{cases}\label{Jm}
\end{align}   
defines an interface spin conductance tensor measuring the electron-spin-to-magnon conversion efficiency through $\mathbf{J}_{s}(x=0)=\overset{\leftrightarrow}{G}\cdot\boldsymbol{\mu}_{s}$:
\begin{align}
\overset{\leftrightarrow}{G}=&\frac{\pi V_{sd}^2\mathcal{N}_{F}^2}{\hbar A_{I}}\text{Im} \sum_{\mathbf{q}\nu}\left\{ s_{\nu} \omega_{\mathbf{q}\nu} \frac{\partial n_{\mathbf{q}\nu}}{\partial\omega_{\mathbf{q}\nu}}\hat{\mathbf{m}}\otimes(\boldsymbol{\xi}_{\nu}^{\ast}\times\boldsymbol{\xi}_{\nu})\right.\nonumber\\
&\left.+ s_{\alpha\beta} \omega_{\mathbf{q}\nu}\frac{\partial n_{\mathbf{q}\nu}}{\partial\omega_{\mathbf{q}\nu} }\hat{\mathbf{n}}\otimes(\boldsymbol{\xi}_{\alpha}^{\ast}\times\boldsymbol{\xi}_{\beta})\right\}\label{spininje}
\end{align}   
where $\overset{\leftrightarrow}{G}=G_{mm}\hat{\mathbf{m}}\otimes \hat{\mathbf{m}}+G_{mn}\hat{\mathbf{m}}\otimes \hat{\mathbf{n}}+G_{nm}\hat{\mathbf{n}}\otimes \hat{\mathbf{m}}+G_{nn}\hat{\mathbf{n}}\otimes \hat{\mathbf{n}}$. The components \( G_{mm} \propto \sin^2 \theta \), \( G_{nm(mn)} \propto \varrho_u \sin \theta \cos \theta \), and \( G_{nn} \propto \cos^2 \theta \) depend on the canting angle \( \theta \).  $G_{nm}$ ($G_{mn}$) describes the cross-polarized magnon spin current along $\hat{\mathbf{n}}$ ($\hat{\mathbf{m}}$) injected by a
$\boldsymbol{{\mu}_{s}} \parallel \hat{\mathbf{m}}$ ($\hat{\mathbf{n}}$), which vanishes for collinear AFMs and for compensated interfaces~($\varrho_{u}= 0$). When $\theta$ is small, $G_{mm}\propto\theta^{2}$ and $G_{nm}\propto\theta$ indicate that spin injection by $\boldsymbol{\mu}_{s}\parallel\hat{\mathbf{m}}$ is much less efficient than by $\boldsymbol{\mu}_{s}\parallel\hat{\mathbf{n}}$ since $G_{nn}\propto 1$. This result agrees with the reported vanishing (large) non-local spin transport signal in $\alpha$-Fe$_2$O$_3$ when $\boldsymbol{\mu}_{s}\perp\hat{\mathbf{n}}$ ($\boldsymbol{\mu}_{s}\parallel\hat{\mathbf{n}}$) \cite{lebrun2018tunable}, and can be understood as a consequence of the spin-transfer torque. Moreover, we expect efficient spin injection when $\boldsymbol{\mu}_{s}\parallel \hat{
\mathbf{m}}$ at larger $\theta$, as observed in the spin-canted state of weakly coupled van der Waals antiferromagnets \cite{PhysRevB.107.L180403}. A $\boldsymbol{\mu}_{s}$ perpendicular to both $\hat{\mathbf{n}}$ and $\hat{\mathbf{m}}$ maximizes the spin-transfer torque since no spin current escapes into the nAFM. 

Eq.~(\ref{Sm}) and Eq.~(\ref{Jm}) predict several unique features of magnonic spin transport in nAFMs. (i) The spin injection induces a magnon density and current with spin polarization in the plane spanned by the $\hat{\mathbf{m}}$ and $\hat{\mathbf{n}}$. (ii) The $\hat{\mathbf{n}}$-component decays in an oscillatory manner with a $q$-dependent wavelength $\sim 1/k_{s}^{(\prime)}$, while the $\hat{\mathbf{m}}$-component decays monotonically. (iii) The phase $\phi_{I}$ from the injector shifts the maximum $\hat{\mathbf{n}}$-component signal away from the band crossing or ``degeneracy" point at $\omega_{s}=0$. (iv) When $k_{s}\lambda_{\alpha\beta}\gg 1$, the oscillations in the $\hat{\mathbf{n}}$-components of the spin density and current are phase-shifted by \(\pi/4\). (v) While the magnon Hanle effect in collinear AFMs requires an easy-plane anisotropy~\cite{PhysRevB.102.174445,PhysRevB.107.184404, PhysRevB.110.L140408}, coherent magnon spin oscillations may emerge in many nAFMs with non-degenerate magnon bands.

Fig.~\ref{Fig-transport} presents numerical results for the normalized $\hat{\mathbf{n}}$-component magnon spin accumulation ($\delta S_{n}$) injected by a spin accumulation $\boldsymbol{\mu}_{s}\parallel\hat{\mathbf{n}}$, for both spin-flop and spin-canted states of easy-axis Cr$_2$O$_3$ at room temperature, with $\varrho_{u}=0$, $\tau_{D}=10\,$ns, and $\tau_{0}=0.1\,$ps. Our choice of $D_{\alpha\beta}=10^{-5}\,$m$^2$/s corresponds to a nominal magnon diffusion length $\lambda=\sqrt{D_{\alpha\beta}\tau_{D}}=0.3\,\mu\text{m}$. For a compensated interface, $\delta S_{m}$ and $\phi_{I}$ vanish when $\boldsymbol{\mu}_{s}\parallel \hat{\mathbf{n}}$. $\delta S_{n}$ oscillates as a function of distance ($d$) from the interface and applied magnetic field with a reduced diffusion length. In the spin-flop state, $\delta S_{n}$ peaks at a magnetic field that depends on $d$, whereas in the spin-canted state, an applied field monotonously suppresses $\delta S_{n}$. Cr$_2$O$_3$ differs from the canted easy-plane $\alpha$-Fe$_2$O$_3$ above the Morin transition in that its magnon bands are degenerate at zero magnetic fields. Therefore the magnon Hanle effect observed in the latter material is maximal at all distances for the high compensation magnetic field at which its two magnon bands cross \cite{PhysRevLett.125.247204}. 

\emph{Conclusion.---}We report coherent magnon injection and transport in two-sublattice nAFMs. In contrast to collinear AFMs, electrically injected magnon spin accumulations and currents have polarization components along both the N\'eel vector and the net magnetization. The former undergoes field-controlled oscillations around the non-collinear magnetization texture with unique phase shifts absent in collinear AFMs. We illustrate our findings by predicting a magnon Hanle effect in the spin-flop and spin-canted states of the easy-axis AFM Cr$_2$O$_3$, which behaves distinctly from that observed in the commonly used easy-plane $\alpha$-Fe$_2$O$_3$. While we focus on two-sublattice nAFMs driven by applied
fields, the coherent oscillation of magnon spins by the superposition of non-degenerate magnon bands reflects the non-collinear order of the ground-state magnetization. Therefore, the physical principle for spin injection and transport of coherent magnon superposition states should hold for nAFMs with intrinsic non-collinear spin order, such as Mn$_3$Sn, offering the possibility of developing functional coherent-magnon devices based on a ubiquitous class of materials.

\emph{Acknowledgments.---}P.T. and G.E.W.B. were supported by JSPS KAKENHI Grants (Nos. 22H04965, 19H00645 and 24H02231). P. T. was also supported by JSPS KAKENHI Grant for Early-Career Scientists (No. 23K13050).


\bibliography{reference}

\begin{thebibliography}{34}
\bibitem [{Not()}]{Notevarrho}%
  \BibitemOpen
  \href@noop {} {}\bibinfo {note} {{$\varrho_{u}$ is defined in the same manner as that of the N\'eel vector $\mathbf{n}=(\mathbf{S}_{a}-\mathbf{S}_{b})$, such that $\varrho_{u}\hat{\mathbf{n}}$ follows the equilibrium orientation of the sublattice spin with dominant s-d exchange interaction.}}\BibitemShut {Stop}%
\bibitem [{\citenamefont {{To leading order in the interfacial s-d exchange coupling we disregard here the backflow spin current arising from a magnon accumulation at the interface}}()}]{NoteN}%
  \BibitemOpen
  \bibfield  {author} {\bibinfo {author} {\bibnamefont {{To leading order in the interfacial s-d exchange coupling we disregard here the backflow spin current arising from a magnon accumulation at the interface}}},\ }\href@noop {} {}\bibinfo {note} {{which is allowed when the magnon conductivity is sufficiently large.}}\BibitemShut {Stop}%
\end{thebibliography}
\end{document}